\documentclass[a4paper,fleqn,usenatbib]{mnras}

\usepackage[T1]{fontenc}
\usepackage{ae,aecompl}

\usepackage{graphicx}	
\usepackage{amsmath}	
\usepackage{amssymb}	
\usepackage{breakurl}
\usepackage{pdflscape}

\newcommand{\ditto}{$- \prime \prime -$}
\newcommand{\dg}{^{\circ}}

\title[RoboPol: EVPA rotations in blazars]{RoboPol: Do optical polarization rotations occur in all blazars?}

\author[D. Blinov et al.]
{D. Blinov$^{1,2,3}$\thanks{E-mail: blinov@physics.uoc.gr}, V. Pavlidou$^{1,2}$, I. Papadakis$^{1,2}$, S. Kiehlmann$^{4,5}$,
I. Liodakis$^{1,2}$,
\newauthor
G.\,V. Panopoulou$^{1,2}$, T.\,J. Pearson$^{7}$, E. Angelakis$^{6}$, M. Balokovi\'{c}$^{7}$, T. Hovatta$^{4,5}$,
\newauthor
V. Joshi$^{8}$, O.\,G. King$^{7}$, A. Kus$^{9}$, N. Kylafis$^{2,1}$, A. Mahabal$^{7}$, A. Marecki$^{9}$,
\newauthor
I. Myserlis$^{6}$, E. Paleologou$^{1,2}$, I. Papamastorakis$^{1,2}$, E. Pazderski$^{9}$,  S. Prabhudesai$^{8}$,
\newauthor
A. Ramaprakash$^{8}$, A.\,C.\,S. Readhead$^{7}$, P. Reig$^{2,1}$, K. Tassis$^{1,2}$, J. A. Zensus$^{6}$ \\
$^{1}$Department of Physics and Institute for Plasma Physics, University of Crete, 71003, Heraklion, Greece\\
$^{2}$Foundation for Research and Technology - Hellas, IESL, Voutes, 7110 Heraklion, Greece\\
$^{3}$Astronomical Institute, St. Petersburg State University, Universitetsky pr. 28, Petrodvoretz, 198504 St. Petersburg,
Russia \\
$^{4}$Aalto University Mets\"ahovi Radio Observatory, Mets\"ahovintie 114, 02540 Kylm\"al\"a, Finland \\
$^{5}$Aalto University  Department of Radio Science and Engineering, P.O. BOX 13000, FI-00076 AALTO, Finland \\
$^{6}$Max-Planck-Institut f\"{u}r Radioastronomie, Auf dem H\"{u}gel
69, 53121 Bonn, Germany\\
$^{7}$Cahill Center for Astronomy and Astrophysics, California Institute of Technology, 1200 E California Blvd,
MC 249-17,\\Pasadena CA, 
91125, USA\\
$^{8}$Inter-University Centre for Astronomy and Astrophysics, Post Bag
4, Ganeshkhind, Pune - 411 007, India\\
$^{9}$Toru\'{n} Centre for Astronomy, Nicolaus Copernicus University, Faculty of Physics, Astronomy and Informatics,
Grudziadzka 5,\\ 87-100 Toru\'{n}, Poland
}

\date{Accepted XXX. Received YYY; in original form ZZZ}

\pubyear{2016}

\begin{document}
\label{firstpage}
\pagerange{\pageref{firstpage}--\pageref{lastpage}}
\maketitle

\begin{abstract}
We present a new set  of optical polarization plane rotations in blazars, observed during the third year of operation
of {\em RoboPol}. The entire set of rotation events discovered during three years of observations is analysed with the aim of
determining whether these events are inherent in all blazars. It is found that the frequency of the polarization plane
rotations varies widely among blazars. This variation cannot be explained either by a difference in the relativistic
boosting or by selection effects caused by a difference in the average fractional polarization. We conclude that
the rotations are characteristic of a subset of blazars and that they occur as a consequence of their intrinsic properties.
\end{abstract}




\section{Introduction} \label{sec:introduction}
Blazars are active galactic nuclei with relativistic jets oriented toward the observer. Relativistic boosting
causes synchrotron radiation from the jet  to dominate the blazar spectra at low 
frequencies \citep{Blandford1979}. Consequently, the optical emission of blazars often has high and
variable polarization. Commonly, the polarization fraction and the electric vector position angle (EVPA) in the
optical band show irregular variations \citep[e.g.][]{Brindle1985}. However, a number of events have been detected
in which the EVPA traces continuous, smooth rotations that in some cases occur contemporaneously with flares in the
total broadband emission \citep{Marscher2008}.

It has been suggested that at least some large amplitude EVPA swings can be physically associated with gamma-ray flares
\citep[e.g.][]{Larionov2013,Blinov2015,Zhang2016}. The {\em RoboPol} programme\footnote{\url{http://robopol.org}} has been
designed for efficient detection of EVPA rotations in statistically rigorously defined samples of gamma-ray--loud
and gamma-ray--quiet blazars and to investigate possible correlations between their gamma-ray activity and optical
EVPA variability \citep{Pavlidou2014}.

{\em RoboPol} started observations at Skinakas observatory, Greece, in May 2013. The EVPA rotations detected during its first two years
of operation were presented in \citet[hereafter Papers~I and II]{Blinov2015,Blinov2016}. In Paper~I we presented
evidence that at least some EVPA rotations must be physically connected to the gamma-ray flaring activity. We also found
that the most prominent gamma-ray flares occur simultaneously with EVPA rotations, while fainter flares may
be non-contemporaneous with the rotations. This was taken as evidence for the co-existence of two separate
mechanisms producing the EVPA rotations. In Paper~II we showed that the polarization degree decreases during the EVPA
rotation events. The magnitude of this decrease is related to the rotation rate in the jet reference
frame. Moreover we presented indications that the EVPA rotations cannot be of arbitrary duration and amplitude.

In this paper, we present a new set of EVPA rotations that were detected during the third {\em RoboPol} observing season 
in 2015. Then, using data from all three seasons we study the occurrence frequency of the EVPA rotations in
blazars. We aim to determine whether EVPA rotations occur in all blazars with the same frequency and to investigate
whether the rotation events are related to the activity of the sources in the gamma-ray band.

\section{Observations, data reduction and detected EVPA rotations} \label{sec:observations}

The third {\em RoboPol} observing season started in 2015 May and lasted until the end of 2015 November. During this period we
obtained more than 1200 measurements of objects from our monitoring sample. The sample is composed of three groups:
the main (``gamma-ray--loud'') group of 62 blazars detected by {\em Fermi}-LAT and listed in the 2FGL catalogue \citep{Nolan2012};
a control group of 17 ``gamma-ray--quiet'' blazars; and an additional group of 24 sources of high interest (see \citet{Pavlidou2014}
for details of the sample selection). The control sample originally included 15 sources, but two of
them have been detected by {\em Fermi}-LAT since the start of our project and are listed in the 3FGL catalogue \citep{Acero2015}.
We therefore included these two sources, which had not been detected previously by {\em Fermi}-LAT, in the main sample
for the third observing season.

\subsection{Observations and data reduction} \label{subsec:datared}

All the polarimetric data analysed in this paper were obtained at the 1.3-m telescope of Skinakas
observatory using the {\em RoboPol} polarimeter. The polarimeter was specifically designed for this
monitoring programme. It has no moving parts besides the filter wheel and thus avoids unmeasurable errors caused
by sky changes between measurements and the non-uniform transmission of a rotating optical element. 
The instrument and the specialized pipeline with which the
data were processed are described in \cite{King2014}. 

The data were taken in the $R$-band. Magnitudes were calculated using calibrated field
stars either found in the literature\footnote{https://www.lsw.uni-heidelberg.de/projects/extragalactic/charts/} or
presented in the Palomar Transient Factory catalogue \citep{Ofek2012}. Photometry of
blazars with bright host galaxies was performed with a constant 4 arcsec aperture. All other sources were measured with
an aperture defined as $2.5 \times \text{FWHM}$, where FWHM is the average full width at half maximum of stellar images
in the 13$\times$13 arcmin field and has a median value of 2.1 arcsec.

The exposure time was adjusted according to  the brightness of each target, which was estimated during a short pointing exposure.
Typical exposures for targets in our sample were in the range 2--30 minutes. The average relative photometric error
was $\sim 0.04$\,mag. Objects in our sample have Galactic latitude $|b| > 10\dg$,
so the average colour excess in the directions of our targets is relatively low, $\langle E(B-V)\rangle = 0.11$\,mag
\citep{Schlafly2011}. Consequently, the interstellar polarization is expected to be less than $1.0\%$ on average
\citep{Serkowski1975}. The statistical uncertainty in the measured degree of polarization is less than $1\%$
in most cases, while the EVPA is typically determined with a precision of 1--$10\dg$ depending on the source brightness
and fractional polarization.

We resolve the 180$\dg$ EVPA ambiguity by assuming that the temporal variation is smooth and does not exceed 90$\dg$ between
two consecutive measurements $\theta_{n}$ and $\theta_{{n}+1}$. The variation is defined as $\Delta \theta = |\theta_{{n}+1} - \theta_{n}|$
and considered to be significant if $\Delta \theta > \sqrt{\sigma(\theta_{n+1})^2 + \sigma(\theta_{n})^2}$, where
$\sigma(\theta_{i})$ is the uncertainty of $\theta_{i}$. If $\Delta \theta - \sqrt{\sigma(\theta_{n+1})^2 + \sigma(\theta_{n})^2} > 90\dg$,
we shift the angle $\theta_{{n}+1}$ by $\pm \, k \times 180\dg$, where the integer $\pm \, k$ is chosen in such a way that
it minimizes $\Delta \theta$. Otherwise, we leave $\theta_{{n}+1}$ unchanged.

\begin{figure*}
 \centering
 \includegraphics[width=0.45\textwidth]{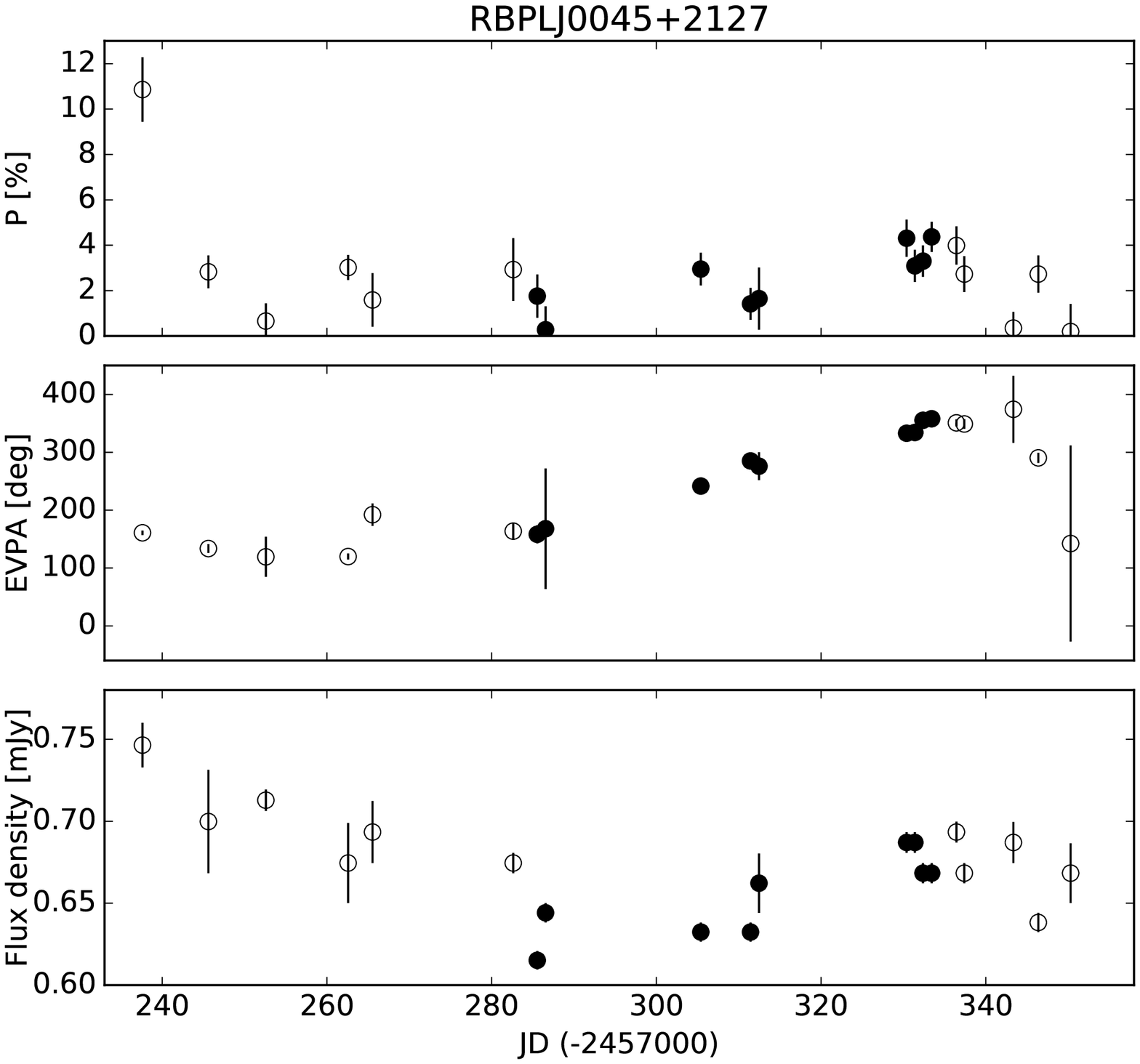}
 \includegraphics[width=0.45\textwidth]{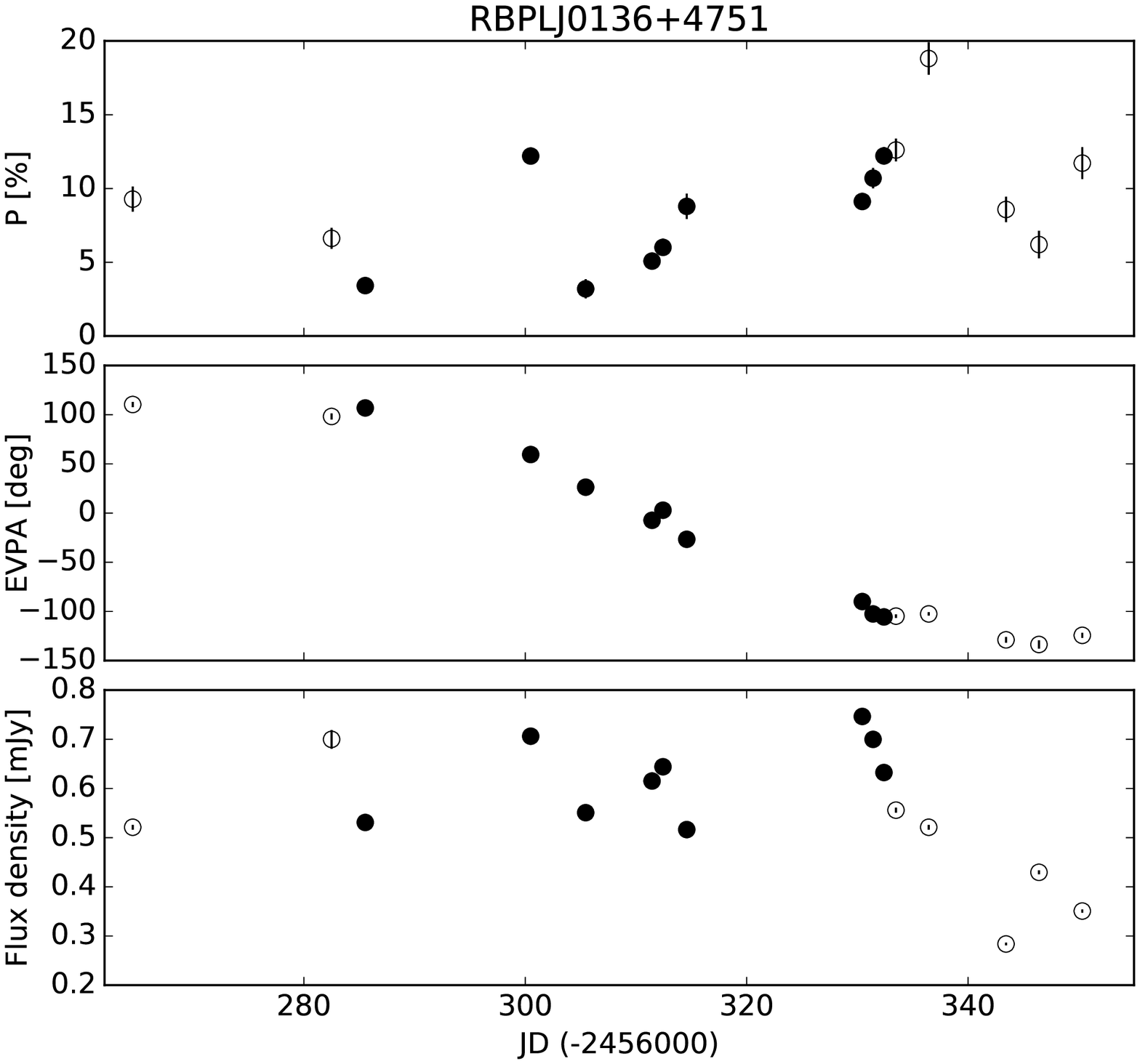}\\
 \includegraphics[width=0.45\textwidth]{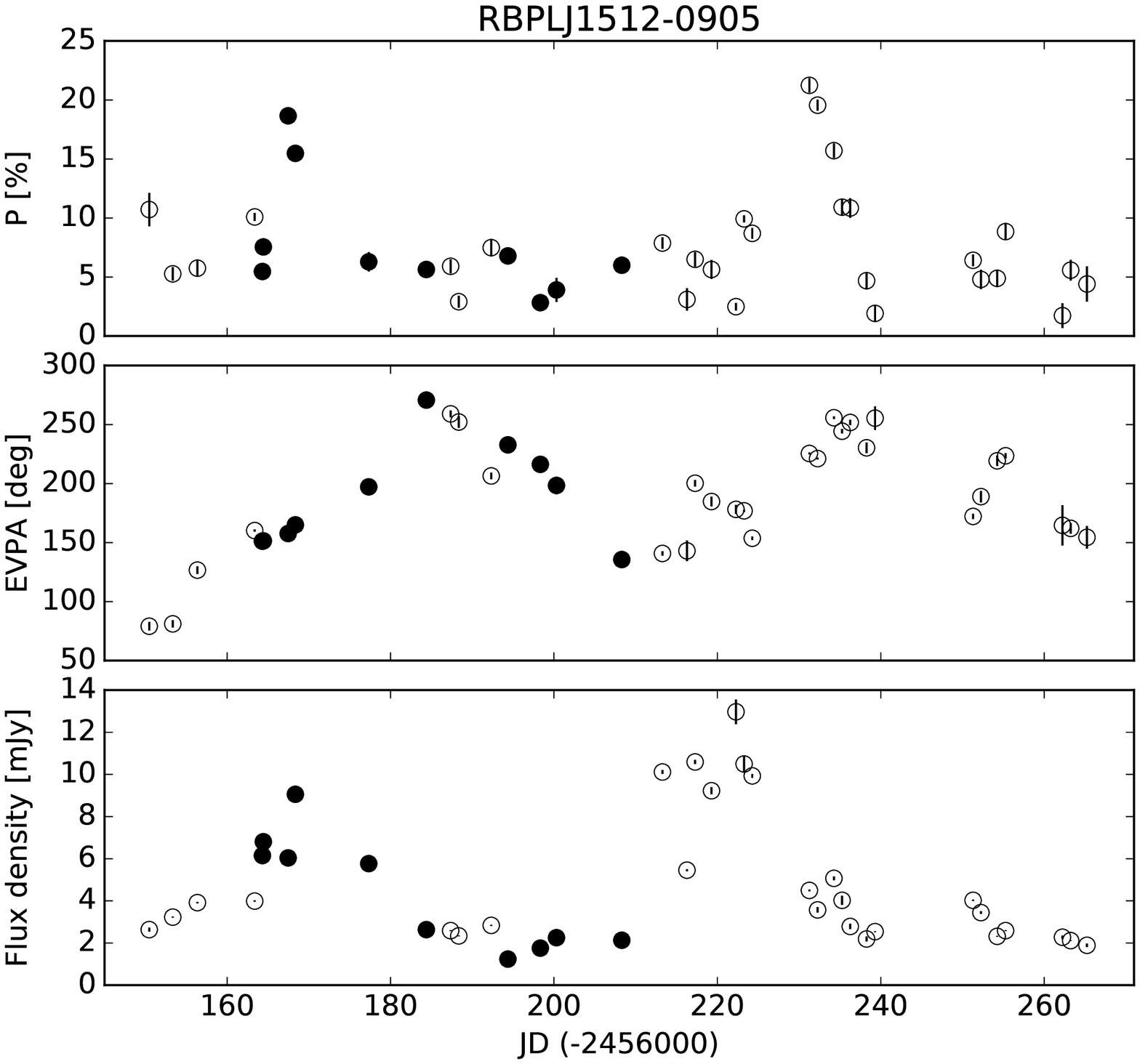}
 \includegraphics[width=0.45\textwidth]{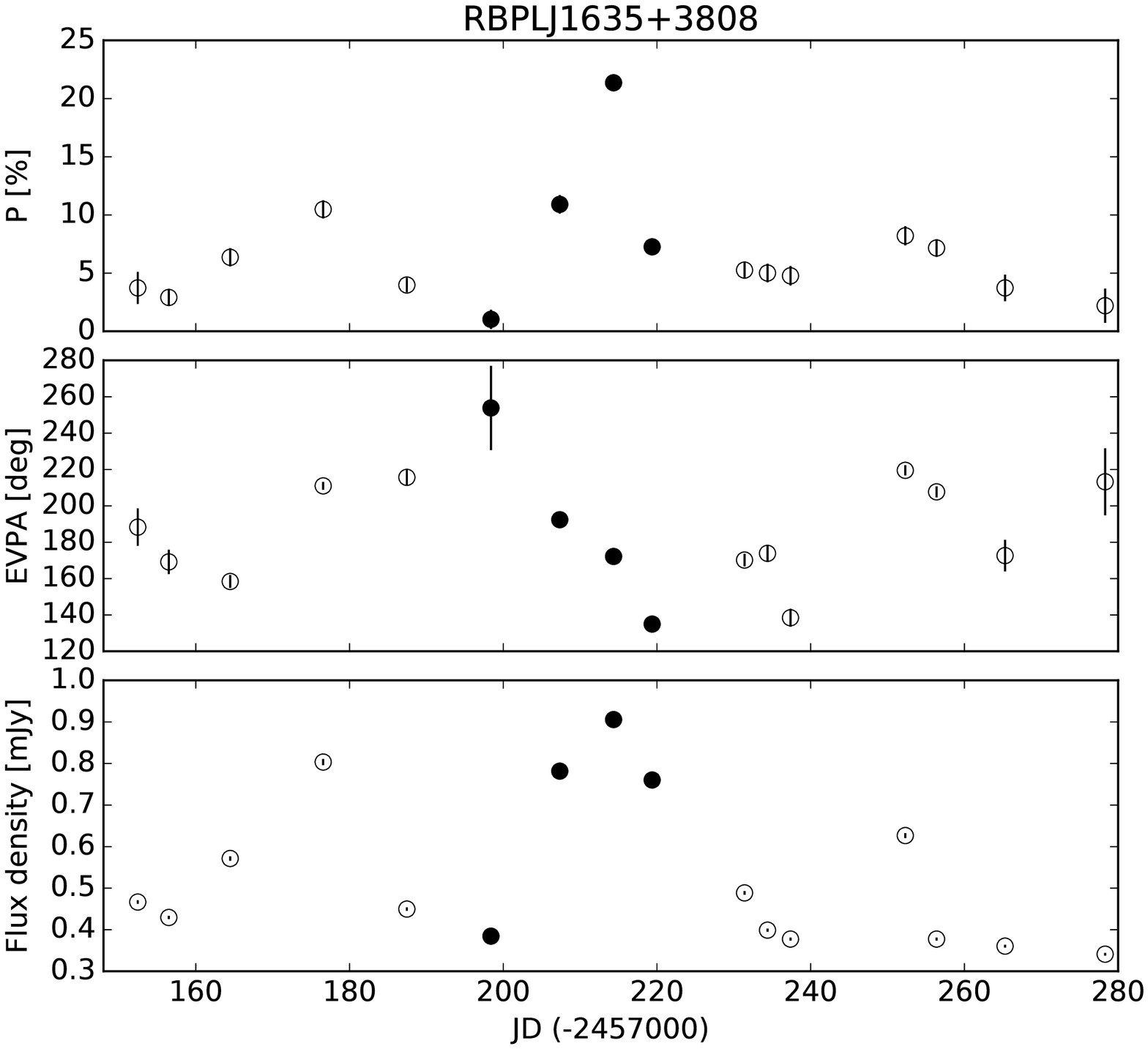}\\
 \includegraphics[width=0.45\textwidth]{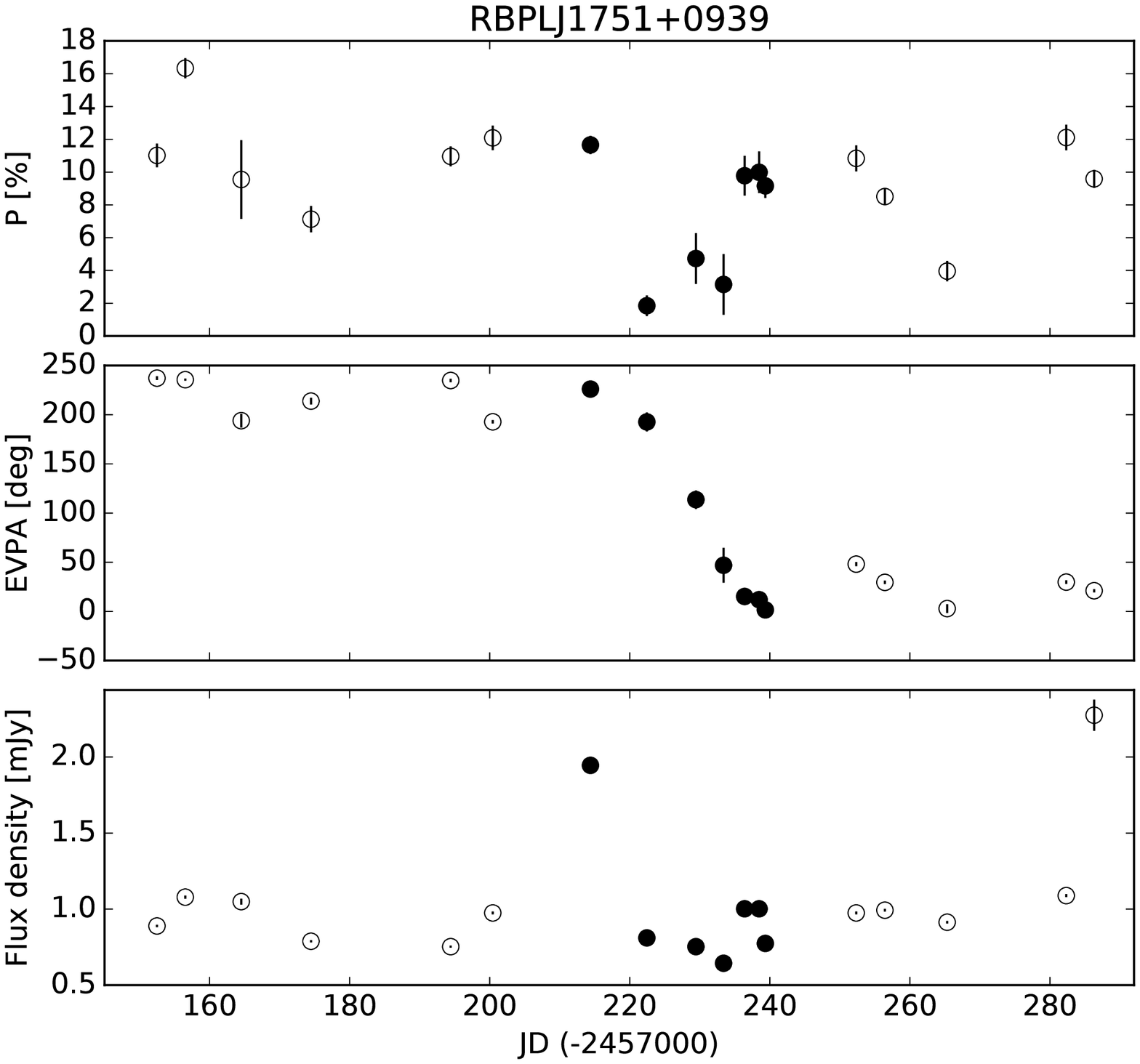}
 \includegraphics[width=0.45\textwidth]{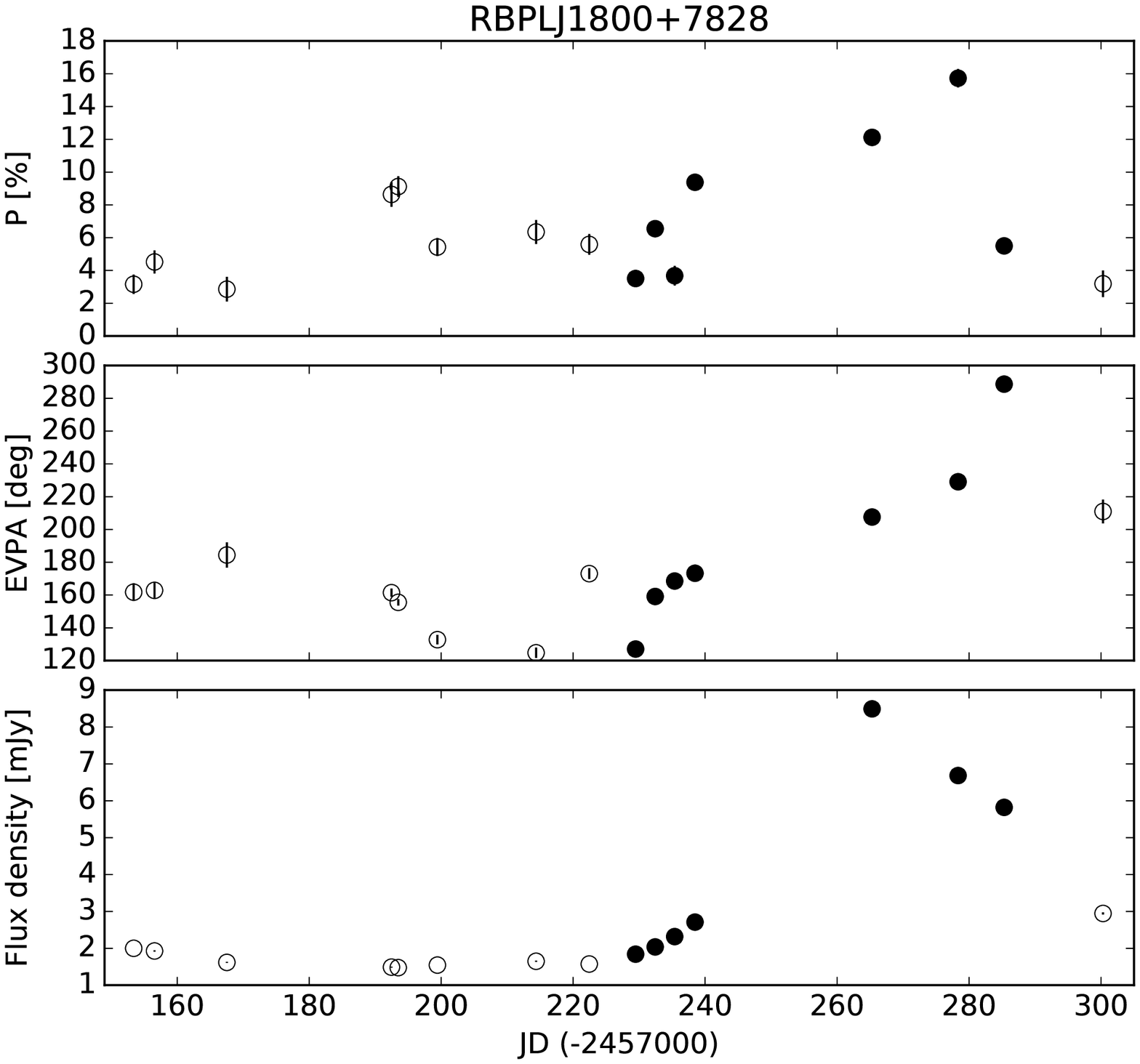}\\
\caption{Evolution of fractional polarization, EVPA and $R$-band flux density for blazars with rotations
detected during the third {\em RoboPol} season. Periods of rotations are marked by filled circles.}
\label{fig:rotations}
\end{figure*}
\begin{figure*}
 \centering
 \includegraphics[width=0.45\textwidth]{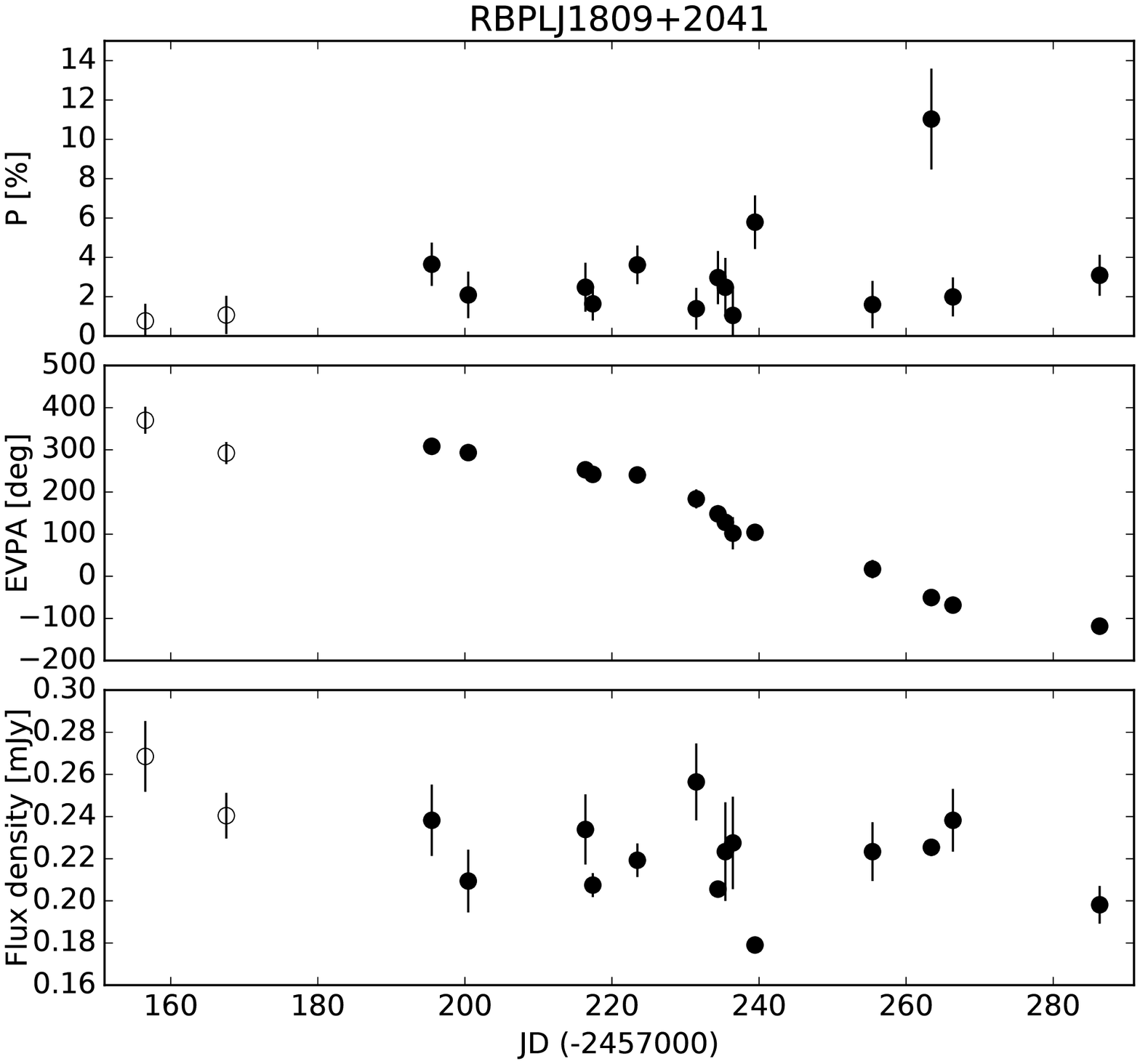}
 \includegraphics[width=0.45\textwidth]{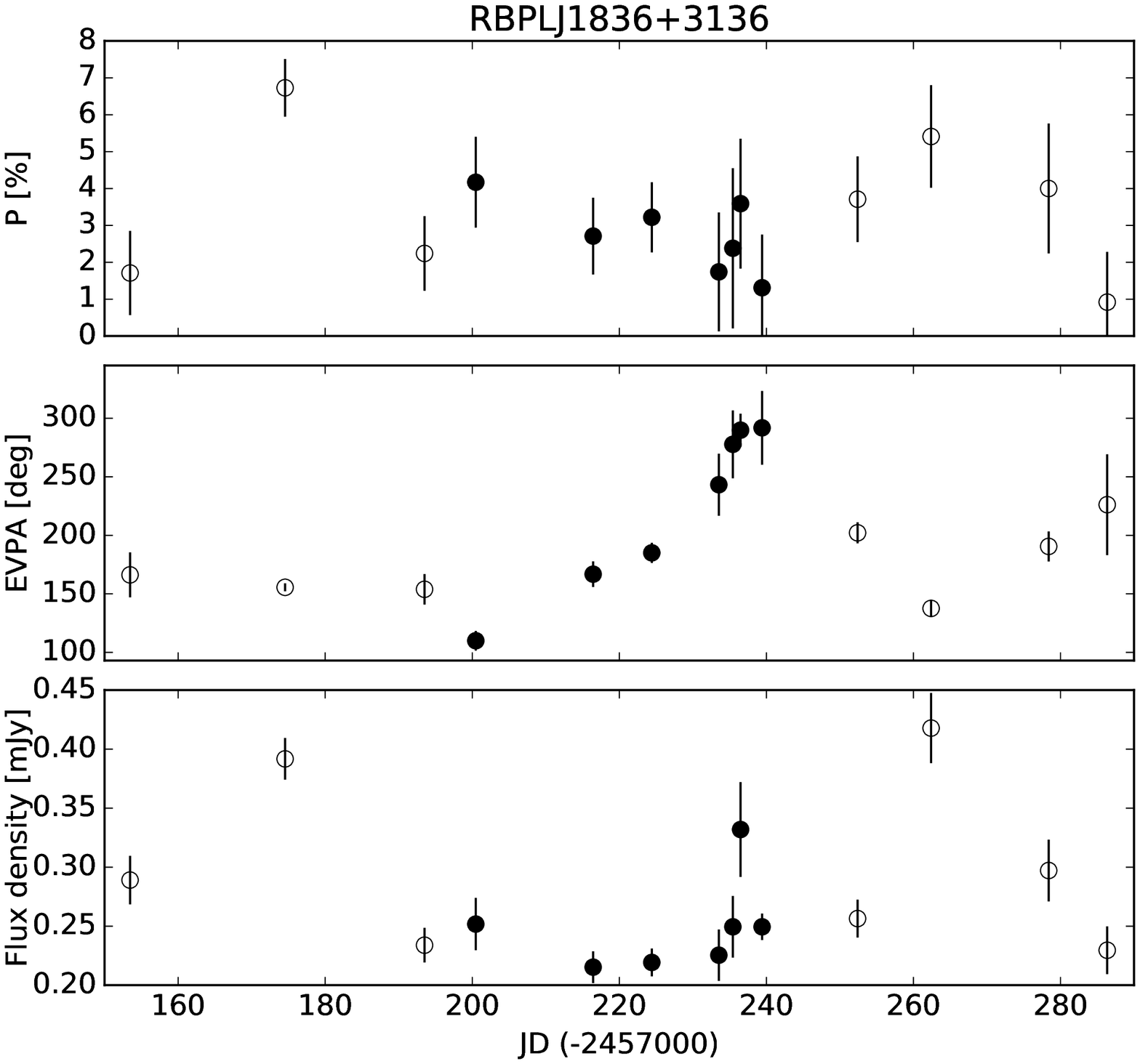}\\
 \includegraphics[width=0.45\textwidth]{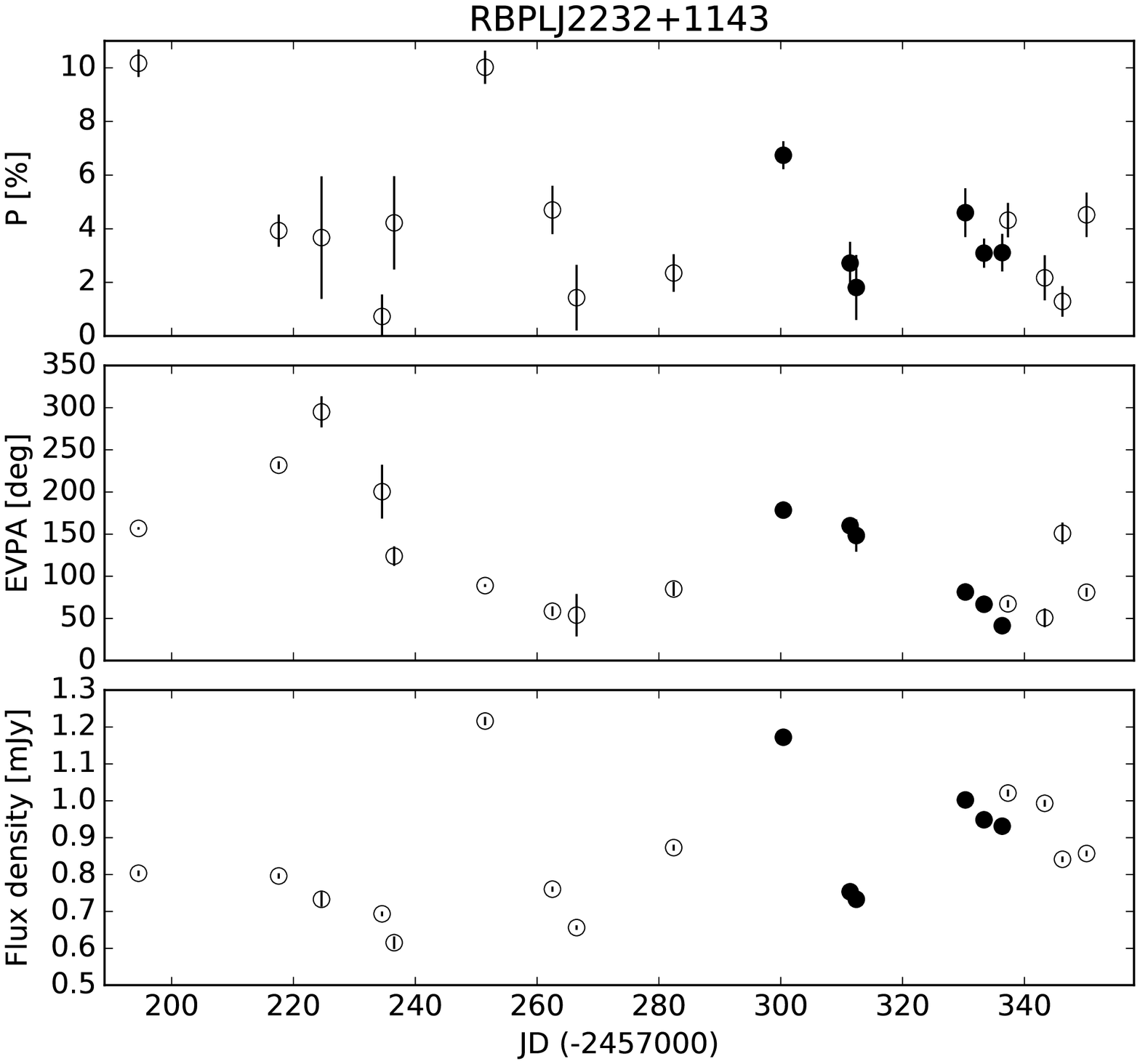}
 \includegraphics[width=0.45\textwidth]{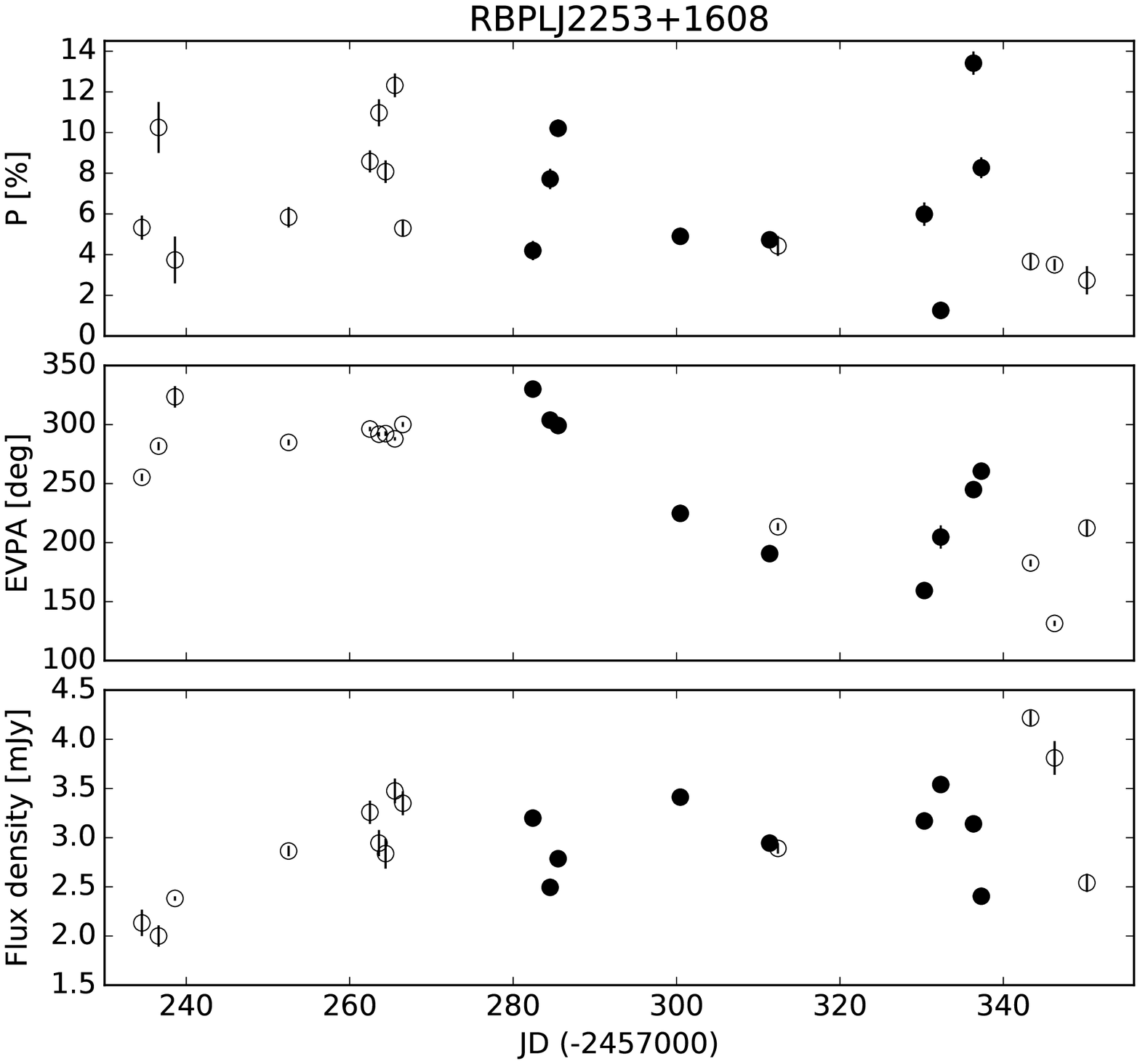}\\
\contcaption{\label{fig:rotations2}}
\end{figure*}

\begin{table*}
\centering
\caption{Observational data for the EVPA rotations detected by {\em RoboPol} in 2015. Columns (1), (2): blazar identifiers;
(3): 2015 observing season length; (4): median time difference between consecutive observations; (5): total amplitude
of EVPA change; (6): duration of the rotation; (7): number of observations during rotation; (8):  average rotation rate.}
\label{tab:rbpl_rotations}
  \begin{tabular}{lccccccc} 
  \hline
Blazar ID      &   Survey         &  $T_{\rm obs}$ & $\langle\Delta t\rangle$ & $\Delta \theta_{\rm max}$ &   $T_{\rm rot}$ & $N_{\rm rot}$ &  $\langle \Delta \theta /\Delta T \rangle$ \\
               &   name           &           (d)  &      (d)                 &        (deg)              &     (d)         &               &             (deg d$^{-1}$)                 \\
 \hline
RBPL\,J0045+2127  & GB6 J0045+2127 &  113   &   3.0  &   199.8      &  48  &  9 &   4.2  \\
RBPL\,J0136+4751  & OC 457         &   86   &   3.0  &  $-$114.2    &  26  &  4 & $-$4.4 \\
RBPL\,J0136+4751  & \ditto         & \ditto & \ditto &  $-$108.5    &  20  &  5 & $-$5.4 \\
RBPL\,J1512$-$0905& PKS 1510$-$089 &  115   &   2.0  &   119.6      &  20  &  6 &   6.0  \\
RBPL\,J1512$-$0905& \ditto         & \ditto & \ditto &  $-$97.2     &  14  &  4 & $-$7.0 \\
RBPL\,J1635+3808  & 4C 38.41       &  126   &   9.0  &  $-$118.9    &  21  &  4 & $-$5.7 \\
RBPL\,J1751+0939  & OT 081         &  134   &   7.0  &  $-$224.5    &  25  &  7 & $-$9.0 \\
RBPL\,J1800+7828  & S5 1803+784    &  147   &   7.5  &   161.6      &  56  &  7 &   2.9  \\
RBPL\,J1809+2041  & RX J1809.3+2041&  130   &   6.0  &  $-$426.7    &  91  & 14 & $-$4.7 \\
RBPL\,J1836+3136  & RX J1836.2+3136&  133   &   9.0  &   181.8      &  39  &  7 &   4.7  \\
RBPL\,J2232+1143  & CTA 102        &  156   &   6.5  &  $-$137.1    &  36  &  6 & $-$3.8 \\
RBPL\,J2253+1608  & 3C 454.3       &  116   &   2.0  &  $-$139.4    &  29  &  5 & $-$4.8 \\
RBPL\,J2253+1608  & \ditto         & \ditto & \ditto &   101.2      &  7   &  4 &  14.5  \\
\hline
\end{tabular}
\end{table*}

\subsection{Detected EVPA rotation events} \label{subsec:detection}

Following Papers I and II, we define an EVPA rotation as any continuous change in the EVPA that is indicated by at
least four consecutive measurements with at least three significant swings between them, and has a total amplitude of
$\Delta \theta_{\rm max} \ge 90\dg$. Moreover, the EVPA curve slope $\Delta \theta_{i}/\Delta t_{i}$ has to change by no more
than a factor of five between consecutive pairs of measurements, and must preserve its sign.

In the data set obtained during the 2015 observing season we identified 13 events in 10 blazars of the main sample that
satisfy our definition of an EVPA rotation. The full season EVPA curves along with the evolution of the polarization
degree and the $R$-band flux density, for the 10 blazars with detected rotations, are shown in Fig.~\ref{fig:rotations}.
The EVPA rotation intervals are marked by the filled circles. The observational parameters of the rotations: the amplitude,
$\Delta \theta_{\rm max}$, and the average rate, $\langle \Delta \theta /\Delta T \rangle$, are listed in Table~\ref{tab:rbpl_rotations},
along with the observing season length, $T_{\rm obs}$, and the median cadence of observations, $\langle\Delta t\rangle$,
for the corresponding blazar. The EVPA swing event in RBPL\,J0136+4751 might be considered to be a single
rotation, but according to our definition it is composed of two successive rotations separated by a significant
swing in the opposite direction to the global trend.

\begin{figure*}
 \centering
 \includegraphics[width=0.99\textwidth]{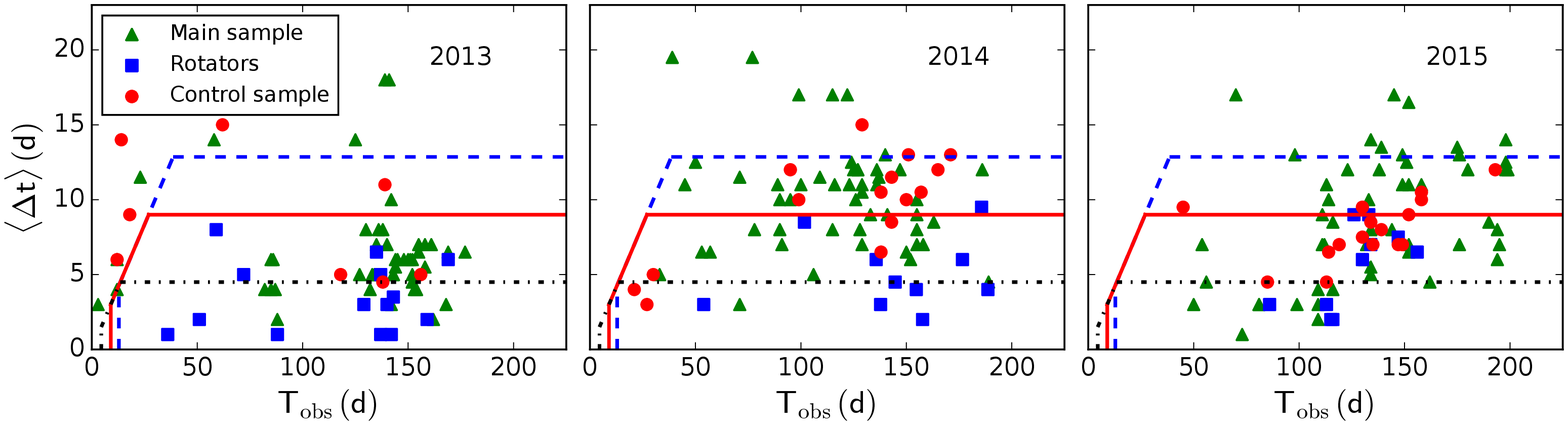}
\caption{Season length, $T_{\rm obs}$, and median cadence, $\langle\Delta t\rangle$, for blazars with detected rotations
for all observing seasons. The lines border areas inside which rotations slower than 7 (dashed blue), 10 (solid red) and
20 deg d$^{-1}$ (dot-dashed black) can be detected (see text for details).}
 \label{fig:cad_len}
\end{figure*}

\section{Frequency of EVPA rotations in blazars} \label{sec:freq}
During the 2013, 2014 and 2015 observing seasons we detected 40 EVPA rotation events in 24 blazars (see Papers I and II for details
of the first two seasons). Two events reported in Paper I belong to neither the main nor the control sample.
Two more events from Paper I do not follow our definition of an EVPA rotation strictly. These four events will not be
taken into account in the analysis below. Using the remaining 36 rotations, in the following sections we address the
question: do {\em all} blazars show EVPA rotations in the optical band?

\subsection{Main and control sample blazars as a single population} \label{subsec:allsame}
A major advantage of the {\em RoboPol} project is that it was operated in such a way that the objects in the two samples would
be observed in a ``similar'' way. Nevertheless, the median observing cadence, $\langle \Delta t \rangle$, and the season
length, $T_{\rm obs}$, are not identical for all the blazars that we monitored. In Fig.~\ref{fig:cad_len} we show $\langle \Delta t \rangle$
versus $T_{\rm obs}$ for each object we observed during each of the observing seasons. The lines in Fig.~\ref{fig:cad_len}
bound regions (``detection boxes'') in the $\langle \Delta t \rangle$ - $T_{\rm obs}$ plane where a rotation slower than
a given rate could have been detected for each object within the area (see Sect. 3.3 of Paper I for details). For example,
the dashed line in Fig.~\ref{fig:cad_len} indicates the maximum $\langle \Delta t \rangle$ value  for a given duration
of observations, $T_{\rm obs}$, that is needed to detect rotations with a rate of
$\langle \Delta \theta/\Delta T \rangle \le 7$ deg d$^{-1}$. We are confident that we can detect rotations with
$\langle \Delta \theta/\Delta T \rangle < 7$ deg d$^{-1}$ for all the blazars within the 7 deg d$^{-1}$ detection box.
The solid and the dash-dotted lines in Fig.~\ref{fig:cad_len} show the respective 10 and 20 deg d$^{-1}$ detection boxes.

In order to compare the EVPA rotation frequencies in blazars that belong to different sub-samples, we need to
consider data from sources in the same detection boxes. The choice of the rotation rate limit is a trade-off between the
number of sources within the detection box and the investigation of a wider range of EVPA rotation rates.
For example, the choice of 7 deg d$^{-1}$ allows us to use data from a number of sources that is substantially
larger than the number of sources that are ``complete'' in the detection of rotations with a rate of $\le 20$ deg d$^{-1}$,
although the latter encompasses a larger portion of all possible EVPA rotation events. In the analysis below, we consider
the objects in all three detection boxes as much as possible.

Columns~1  and~2 in Table~\ref{tab:rots_norots} list the number of detected rotations, $N_{\rm rot}$, slower than 
the given rate and the total observing length, $T_{\rm obs}$, for all blazars within the 7, 10 and 20 deg d$^{-1}$
detection boxes. For instance, the top panel in Table~\ref{tab:rots_norots} takes into account only sources located within
the 7 deg d$^{-1}$ detection boxes for all three seasons, and rotations only with a rate of $\le 7$ deg d$^{-1}$.
Column~3 gives the average frequency of rotations, $\lambda$, slower than the given rate. This
frequency is defined as $\lambda = N_{\rm rot}/T_{\rm obs}$, for $N_{\rm rot}>0$. In the case of $N_{\rm rot}=0$, we
list an upper limit on $\lambda$, which is defined as $1/T_{\rm obs}$. The corresponding numbers are listed separately
for blazars with and without detected EVPA rotations (rotators and non-rotators hereafter).
\begin{table} 
\centering
\caption{Estimates of rotations frequencies. Columns (1) - total number of rotations;
(2) - summed observing length; (3) - average frequency of rotations; (4) - probability of observing $N_{\rm rot}$ during
$T_{\rm obs}$ if all blazars have equal frequency of rotations $\lambda_{\rm all}$.}
\label{tab:rots_norots}
  \begin{tabular}{rcccc} 
  \hline
               & $N_{\rm rot}$     &    $T_{\rm obs}$ (d) &  $\lambda$ (d$^{-1}$) &  $\mathcal{P}$ \\
 \hline
  \multicolumn{5}{c}{$\le$ 7 deg d$^{-1}$}\\
  all:         & 22                &    24584             &  $8.9\times10^{-4}$   &  -                \\
rotators:      & 22                &    6296              & ($3.5\times10^{-3}$)  & $1\times10^{-7}$ \\
non-rotators:  & 0                 &    18288             & ($<5.5\times10^{-5}$) & $9\times10^{-8}$  \\
 \hline
  \multicolumn{5}{c}{$\le$ 10 deg d$^{-1}$}\\
  all:         & 24                &    17820             &  $1.4\times10^{-3}$   & - \\
rotators:      & 24                &    5847              & ($4.1\times10^{-3}$)  & $4\times10^{-6}$ \\
non-rotators:  & 0                 &    11973             & ($<8.4\times10^{-5}$) & $5\times10^{-8}$ \\
 \hline
  \multicolumn{5}{c}{$\le$ 20 deg d$^{-1}$}\\
  all:         & 20                &    5412              &  $3.7\times10^{-3}$   &  - \\
rotators:      & 20                &    2224              & ($9.0\times10^{-3}$)  & $2\times10^{-4}$  \\
non-rotators:  & 0                 &    3188              & ($<3.1\times10^{-4}$) & $8\times10^{-6}$  \\
\hline
\end{tabular}
\end{table}
From the table we can see that $\lambda$ differs by more than an order of magnitude between rotators and non-rotators.

The probability that $n$ independent events occur in a period of time $t$ can be estimated using the Poisson distribution,
\begin{equation}\label{eq:1}
 \mathcal{P}(n,t,\lambda) = \frac{(\lambda t)^n}{n!} e^{-\lambda t},
\end{equation}
where $\lambda$ is the average frequency of the events. Then, using data from Table~\ref{tab:rots_norots}, under the
hypothesis that {\em all blazars exhibit rotations with equal frequency}, we can estimate the probability of having $N_{\rm rot}$
rotations in blazars that were observed over a period of time $T_{\rm obs}$. For instance, the probability of having
22 rotations slower than 7\,deg\,d$^{-1}$ in blazars that fall in the corresponding detection boxes in Fig.~\ref{fig:cad_len}
and were observed for 6296\,d is $\mathcal{P}(22,6296,8.9\times10^{-4}) = 1\times10^{-7}$. This result indicates that,
under the hypothesis of the same frequency of rotations in blazars, it is highly unlikely to detect such a large
number of rotations in a small number of objects observed in such a short $T_{\rm obs}$.

Following the same reasoning we found the corresponding probabilities, $\mathcal{P}$, for rotators and non-rotators within the three
detection boxes. These probabilities, presented in Column~4 of Table~\ref{tab:rots_norots}, are less than
$2\times10^{-4}$ in all cases. Therefore, the null hypothesis is rejected at a high significance level for all detectable rotation
rates: {\em it is highly unlikely that all blazars exhibit rotations of the polarization plane with rates}
$\le 20$\,deg\,d$^{-1}${\em, with the same frequency}.

\subsection{Absence of rotations in the control sample} \label{subsec:contr}

If EVPA rotations are related to the gamma-ray activity of blazars, the low probabilities we found in the previous
section may perhaps arise from the fact that we considered both gamma-ray--loud and gamma-ray--quiet sources as a single population
in the analysis above. Here we test whether the two classes of blazars differ significantly in the frequency of EVPA
rotation they exhibit.

As shown in the previous section, 22 rotations occurred in blazars that are located within the 7\,deg\,d$^{-1}$
detection boxes and have rates slower than 7\,deg\,d$^{-1}$. The total observing length for the main sample blazars in the
7\,deg\,d$^{-1}$ detection boxes for all three seasons is 20625\,d. Thus we can estimate the frequency of rotations with
$\langle \Delta \theta/\Delta T \rangle < 7$\,deg\,d$^{-1}$ in the main sample sources as one rotation in $\sim 940$\,d
($T_{\rm obs}=20625\,\textrm{d} / N_{\rm rot}= 22$ rotations). Following the same rationale we estimate average frequencies of
rotations with rates $ \le 10$\,deg\,d$^{-1}$ and $ \le 20$\,deg\,d$^{-1}$ as one rotation in $\sim 650$\,d ($15632/24$) and
$\sim 250$\,d ($5028/20$).

The total $T_{\rm obs}$ for the control sample blazars lying within the 7\,deg\,d$^{-1}$ detection boxes in Fig.~\ref{fig:cad_len}
is 3959\,d. Under the hypothesis that {\em blazars of the control sample show EVPA rotations with the same frequency as
the main sample sources}, we can estimate the probability of not detecting any rotation with
$\langle \Delta \theta/\Delta T \rangle < 7$\,deg\,d$^{-1}$ in the control sample blazars, as $\mathcal{P}(0,3959,1/940) = 1.5\%$.
Similarly, the probability of not detecting rotations slower than 10 and 20\,deg\,d$^{-1}$ in the control sample is 3.5\%
and 22\%, respectively. These numbers imply that we cannot reject the null hypothesis at a sensibly significant level;
it is possible that $\lambda$ is the same for the blazars in the main and the control samples.

These results indicate that the highly significant difference in the frequency of the EVPA rotations in the rotators
and non-rotators is not due to the fact that we did not observe any rotations in the control sample blazars.
The majority of the main sample blazars did not show any rotations either. It is possible then that the rate of EVPA
rotations is not constant even among the blazars of the main sample. We investigate this possibility in the following
section.

\subsection{Subclasses of blazars within the main sample} \label{subsec:rot_subcl}
\begin{table}
\centering
\caption{Frequencies of EVPA rotations in the main sample sources within the 7 deg d$^{-1}$ detection box. Columns
(1) - total number of rotations; (2) - summed observing period length; (3) - average frequency of rotations.}
\label{tab:rots}
  \begin{tabular}{lccc} 
  \hline
               & $N_{\rm rot}$   & $T_{\rm obs}$ (d) &  $\lambda$ (d$^{-1}$)\\
 \hline
 0 rotations   & 0               & 12978    &  $<7.7\times10^{-5}$ \\
 1 rotation    & 12              & 4462     &  $2.7\times10^{-3}$  \\
 2 rotations   & 6               & 1246     &  $4.8\times10^{-3}$  \\
 3 rotations   & 6               & 885      &  $6.8\times10^{-3}$  \\
 4 rotations   & 12              & 1054     &  $1.1\times10^{-2}$  \\
all rotators:  & 36              & 7647     &  $4.7\times10^{-3}$  \\
\hline
total:         & 36              & 20625    &  $1.8\times10^{-3}$  \\
\hline
\end{tabular}
\end{table}
In this section we ascertain whether the occurrence of EVPA rotations in the blazars of the main sample is consistent with
a single population of sources that exhibit rotations with equal frequency. For this purpose we separate the main sample
into five sub-samples: blazars that did not show any rotation, and blazars that had one to four rotations during the whole
observing period. Then, for each group, we count the total number of rotations, $N_{\rm rot}$, the total observing period
length, $T_{\rm obs}$, and the average frequency of rotations, $\lambda$, as defined earlier, using the sources within the
7\,deg\,d$^{-1}$ detection boxes in Fig.~\ref{fig:cad_len}. These data are presented in Table~\ref{tab:rots}. The average
frequency of rotations is strongly non-uniform among the sub-samples. The frequency of rotations $\lambda$ in non-rotators of the
main sample is more than two orders of magnitude smaller than $\lambda$ for the blazars that exhibited four events.

A number of hypotheses can be considered to verify whether the difference of $\lambda$ for these sub-groups of blazars 
is accidental or not. Under the null-hypothesis that {\em all blazars of the main sample represent
a single class and exhibit EVPA rotations with the average frequency of} $\lambda=1.8\times10^{-3}$\,d$^{-1}$, using
Equation~\ref{eq:1} we find the probability of detecting zero rotations in 41 sources observed for 12978\,d is
$\mathcal{P}(0,12978,1.8\times10^{-3})=7.2\times 10^{-11}$. On the other hand, if we assume that {\em all blazars have the same
frequency of rotations equal to} $\lambda=7.7\times10^{-5}$\,d$^{-1}$ (this is the upper limit of the frequency of rotations
for the non-rotators of the main sample), then the probability of detecting 36 events in the whole sample during 20625\,d of
observations is $\mathcal{P}(36,20625,7.7\times10^{-5})=9\times10^{-36}$. Moreover, assuming that the average frequency of
rotations $\lambda=4.7\times10^{-3}$\,d$^{-1}$ is characteristic of all rotators, we find the probability of detecting
12 rotations in the group of 3 blazars each of those exhibited 4 rotations as $\mathcal{P}(12,1054,4.7\times10^{-3})=3.2 \times 10^{-3}$.
We therefore conclude that {\em the frequency of the EVPA rotations is significantly different among blazars of the main sample}.

The analysis above implies that there is a sub-class of objects that exhibit EVPA rotations much more
frequently than others. This difference does not simply depend on whether or not a blazar is detected by {\em Fermi}-LAT.
Even in objects that are in our gamma-ray--loud sample and did not show any rotation, the frequency of the EVPA rotations
must be significantly smaller than the frequency exhibited by the rotators in our sample.

\section{Possible reasons of different observed rotation frequencies} \label{sec:reasons}

The analysis presented above uses data from objects with the same sampling properties in their light curves.
However, there are two more observational factors which may bias our results and conclusions. The first one depends
on possible differences in the accuracy with which we can measure EVPA in rotators and non-rotators. The second is
related to possible intrinsic differences in the rest-frame properties, namely redshift, $z$, and Doppler factor, $\delta$,
of rotators and non-rotators. We address both issues below.

\subsection{Differences in accuracy of EVPA measurements} \label{subsec:lowpol}
It has been shown by \cite{Pavlidou2014} and Angelakis et al. (in prep.) that the control sample blazars are on average 
significantly less polarized than the main sample blazars. This could potentially be the reason for the absence
of EVPA rotation detections in the control sample. Our definition of an EVPA rotation requires three or more
significant swings between four or more consecutive EVPA measurements. Lower fractional polarization leads to higher
uncertainties in the EVPA, so larger errors may hide significant swings. If
non-rotators are significantly less polarized than rotators we might have missed rotations in their
EVPA curves because of this observational bias.

\begin{figure}
 \centering
 \includegraphics[width=0.44\textwidth]{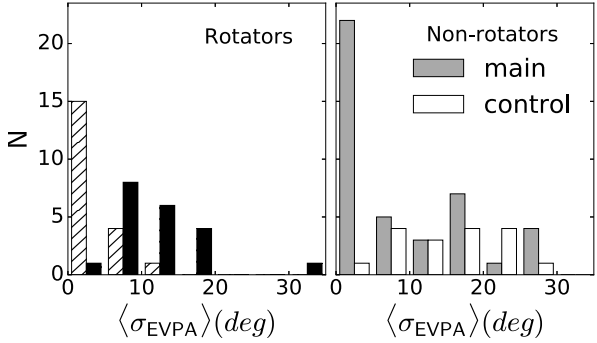}
\caption{Distribution of the median measured uncertainty of the EVPA measurements. Left panel: uncertainties in the rotator 
sub-sample, measured (hatched bars), and multiplied by a factor of four (black bars). Right panel: uncertainties in the 
non-rotators of the main sample and the control sample.}
 \label{fig:med_EVPAerr}
\end{figure}

One way to address this issue is to compare the mean polarization fraction of rotators and non-rotators. Here we choose
to investigate this issue directly, i.e., we compare the amplitude of the EVPA error in rotators and non-rotators. The
distribution of the median uncertainty, $\langle \sigma_{\rm EVPA} \rangle$, of the EVPA measurements for the rotators
and not-rotators is shown in the left and the right panels of Fig.~\ref{fig:med_EVPAerr}. The hatched, grey and white bars
correspond to rotators, non-rotators of the main sample, and non-rotators of the control sample. On average, $\langle \sigma_{\rm EVPA} \rangle$
of non-rotators is larger than $\langle \sigma_{\rm EVPA} \rangle$ of rotators. A number of non-rotators in the main sample
show small uncertainties, but all the control sample sources and quite a few of the main sample non-rotators show large
uncertainties. The two-sample Kolmogorov-Smirnov (K-S) test rejects the hypothesis that the $\langle \sigma_{\rm EVPA} \rangle$
distribution of rotators and non-rotators (all together) is sampled from the same parent population ($p\text{-value}= 3 \times 10^{-4}$).

However, we do not believe that this difference is the main reason why we do not detect rotations in the non-rotators.
To demonstrate this, we multiplied $\sigma_{\rm EVPA}$ of each measurement in the EVPA curves of rotators by
a factor $f$. The distribution of $\langle \sigma_{\rm EVPA} \rangle$ for the rotators when $f=3$ is shown by the black bars
in the left panel of Fig.~\ref{fig:med_EVPAerr}. The null hypothesis that this distribution and the distribution of
$\langle \sigma_{\rm EVPA} \rangle$ for the control sample are drawn from the same parent population cannot be rejected according
to the K-S test ($p\text{-value}= 0.03$). More than half of the rotations (22 out of 36) still follow our definition
of an EVPA rotation for $f=3$.

In order to investigate the significance of the measurement accuracy, we repeated the analysis performed in Sect.~\ref{subsec:allsame}
for rotators and non-rotators within the 7\,deg\,d$^{-1}$ detection box, ignoring the 14 events that do not follow
our definition of an EVPA rotation when $f=3$. Among the remaining rotations that follow the definition, 13 events have
rates slower than 7\,deg\,d$^{-1}$. Therefore, the frequency of rotations decreases in this case, from $\lambda=22/24584=8.9\times10^{-4}$
d$^{-1}$, to $\lambda=13/24584=5.3\times10^{-4}$\,d$^{-1}$. However, even in this case, according to equation~\ref{eq:1},
the probability of detecting 13 rotations in blazars observed for 3589\,d is $\mathcal{P}(13,3589,5.3\times10^{-4})=10^{-7}$.
Similarly, the probability of not detecting any rotations in the remaining blazars observed for 20995\,d is
$\mathcal{P}(0,20995,5.3\times10^{-4})=2\times10^{-5}$. Thus even when we artificially increase the uncertainty of EVPA
measurements in rotators in such a way that $\langle \sigma_{\rm EVPA} \rangle$ distributions for rotators and non-rotators
become comparable, the frequency of rotations cannot be the same for the the two groups.

Therefore the difference in the amplitudes of the EVPA uncertainties could partially explain the absence of detected
rotations in non-rotators. However, it is not large enough to be entirely responsible for the difference in the frequencies
of EVPA rotations between rotators and non-rotators found in Sect.~\ref{sec:freq}.

\subsection{Rest frame timescale differences} \label{subsec:tfactor}
Another possible explanation for the variation in rotation frequency is that we miss rotations in some blazars because
the duration of observations in the jet rest frame, $T_{\rm obs}^{\rm jet}$, may be significantly different for the rotators
and the non-rotators. The analysis in Sect.~\ref{sec:freq} is based on the total number of observing days in the observer
frame, $T_{\rm obs}$, for the rotators and non-rotators. The jet frame and the observer frame timescales are related as
$\Delta T^{\rm jet} = \Delta T^{\rm obs} \delta/(1+z)$. Therefore, $T_{\rm obs}^{\rm jet}$ depends on the Doppler factor,
$\delta$, and the redshift, $z$, of the sources as well. If the $\delta$ and/or $z$ distributions are significantly
different for rotators and non-rotators then the difference in $\lambda$ found above could be artificial.

\begin{figure}
 \centering
 \includegraphics[width=0.43\textwidth]{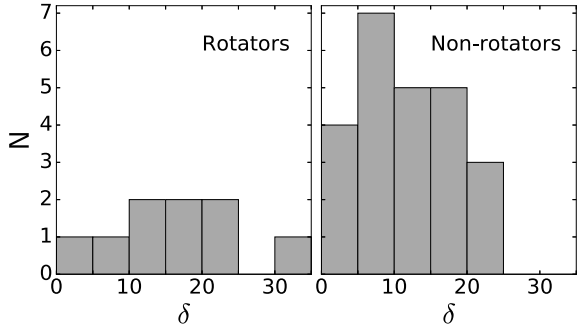}
\caption{Distribution of $\delta$ for rotators and non-rotators.}
 \label{fig:dopp_dist}
\end{figure}

Table~\ref{tab:app} lists estimates of $\delta$ and $z$ for the blazars in our sample taken from the literature.
We use $\delta$ values estimated from the variability of the total flux density in the radio band, which are believed
to be the most reliable and self-consistent Doppler factor estimates available\footnote{The actual Doppler factors for
the optical emission region may be significantly different, but estimates are not available at the moment.}
\citep{Liodakis2015}. Such estimates are available for 21 sources in both samples. Figure~\ref{fig:dopp_dist} shows the
distribution of these $\delta$ values for rotators and non-rotators.
The null hypothesis that the two distributions are drawn from the same population is strongly supported by the data
according to the K-S test ($p\text{-value}= 0.62$).

Redshift estimates are available for 71 sources in the two samples;  52 are based on optical spectroscopic data,
while the other 19 are obtained using indirect methods (broadband photometry of the host galaxies, attenuation of the hard
gamma-ray emission, etc.). According to the K-S test the null hypothesis that the distributions of $z$
for rotators
\begin{figure}
 \centering
 \includegraphics[width=0.44\textwidth]{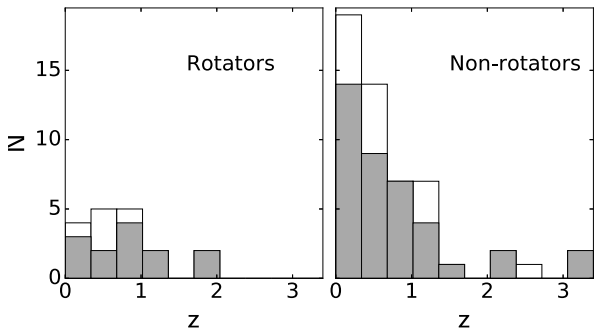}
\caption{Distribution of $z$ for rotators and non-rotators. Grey bars include only spectroscopic $z$ estimates, white
bars include indirect $z$ estimates.}
 \label{fig:z_dist}
\end{figure}
and non-rotators (Fig.~\ref{fig:z_dist})  are drawn from the same population cannot be rejected ($p\text{-value}= 0.48$ for the spectroscopic,
and 0.54 for all available redshifts).

\begin{figure}
 \centering
 \includegraphics[width=0.44\textwidth]{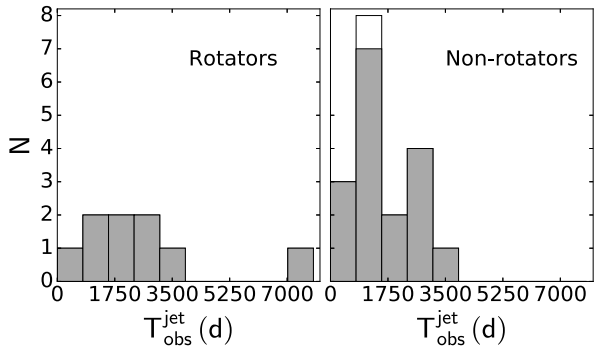}
\caption{Distribution of $T_{\rm obs}^{\rm jet}$ for rotators and non-rotators. Grey bars include only spectroscopic $z$
estimates, white bars include indirect $z$ estimates.}
 \label{fig:tjet_dist}
\end{figure}

For the blazars with known $\delta$ and $z$ that are located in the 7\,deg\,d$^{-1}$ detection boxes in
Fig.~\ref{fig:cad_len}, we computed the total $T_{\rm obs}$ and transformed it to the jet frame $T_{\rm obs}^{\rm jet}$.
The $T_{\rm obs}^{\rm jet}$ distributions for rotators and non-rotators are shown in Fig.~\ref{fig:tjet_dist}.
According to the K-S test, the hypothesis that $T_{\rm obs}^{\rm jet}$ for rotators and non-rotators are drawn from the
same population is again supported by the data ($p\text{-value}= 0.27$ for spectroscopic, and 0.25 for all available redshifts).

Using the reasoning of Sect.~\ref{sec:freq} we can estimate how large the difference between the average Doppler factors of the rotators
and the non-rotators must be to explain the absence of rotations in the EVPA curves of
non-rotators. According to Equation~\ref{eq:1} and the frequency estimate for rotations slower than 7\,deg\,d$^{-1}$ from
Sect.~\ref{subsec:allsame}, if we reduce the total $T_{\rm obs}$ for non-rotators by a factor of 4, then the
probability of having zero rotations in this sub-sample of blazars can reach $\sim$~1\%. This means that if the average
Doppler factor for the non-rotators were 4 times smaller than that of the rotators, then the absence of rotations
could be an accidental outcome of the Poisson distribution of the rotations. Such a large difference is inconsistent
with the aforementioned K-S test that finds no difference in the distributions of $\delta/(1+z)$ for rotators
and non-rotators.

In summary, our analysis strongly suggests that there exists a sub-class of blazars that exhibit EVPA rotations in the optical
band significantly more frequently than the others. The difference in the frequency of the rotations cannot be explained by the
non-uniformity of observations or by observational biases due to differences in the average fractional polarization.
Moreover, based on the available data, it cannot be explained by differences in the relativistic beaming or in the redshifts
between the rotators and non-rotators. In the next two sections we investigate possible physical reasons that might be
responsible for the prevalence of the optical EVPA rotations in this sub-class of blazars.

\section{Rotations in blazars of different classes} \label{sec:rot_classes}

\begin{figure}
 \centering
 \includegraphics[width=0.38\textwidth]{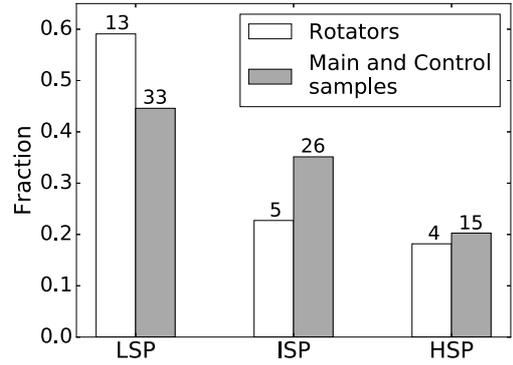}
\caption{Distribution of blazars in the main and control samples together and rotators among the synchrotron peak position
types. Fraction is calculated as the number of sources of a given synchrotron peak class divided by the total
number of sources in the corresponding sample.}
 \label{fig:sync_peak}
\end{figure}

In this section we examine whether the ability of a blazar to exhibit EVPA rotations depends on its synchrotron peak
location. The classification as either a low-, intermediate- or high-synchrotron-peaked (LSP, ISP or HSP) blazar for the
main sample sources was taken from the third catalog of active galactic nuclei detected by the Fermi-LAT \citep[3LAC,][]{Ackermann2015}.
For the control sample sources, which are not in 3LAC, the synchrotron peak positions were taken from Angelakis et al.
(in prep.) and \cite{Mao2016}, where a procedure similar to the one used by \cite{Ackermann2015} was applied. The
classification of blazars in our sample according to the synchrotron peak position is listed in Table~\ref{tab:app}. We
find that the main and the control samples together are composed of 33 LSP, 26 ISP and 15 HSP sources. The sample of
rotators is composed of 13 LSP, 5 ISP and 4 HSP sources. The distribution of the sources among the classes is shown in
Fig.~\ref{fig:sync_peak}. We estimate the probability that rotators comprise sources randomly drawn from the main and
the control samples together as:
\begin{equation}
\mathcal{P} = \frac{C_{33}^{13} C_{26}^{5} C_{15}^{4}}{C_{74}^{22}} = 0.014,
\end{equation}
where $C_{n}^{k}$ is the binomial coefficient. The numerator in this equation is the number of ways to obtain a sample
composed of 13 LSP, 5 ISP and 4 HSP blazars from the parent sample of 33 LSP, 26 ISP and 15 HSP sources. The denominator
is the total number of combinations how 22 objects can be selected out of 74. Similarly, the probability that rotators
are randomly drawn from the main sample only is 0.5\%. Therefore it is unlikely that LSP accidentally dominate
over ISP and HSP among the blazars that exhibit rotations.

\section{Gamma-ray properties of rotators and non-rotators} \label{sec:gamma_prop}
\begin{figure}
 \centering
 \includegraphics[width=0.48\textwidth]{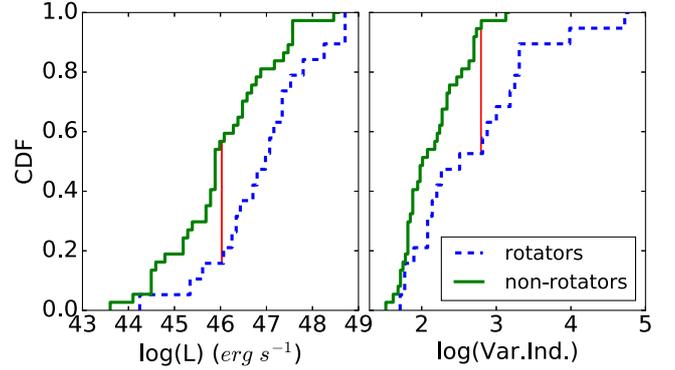}
\caption{CDF of luminosity (left) and variability index (right) for rotators and non-rotators.
The red vertical line indicate the maximum difference between the CDFs.}
 \label{fig:gamma_lum}
\end{figure}

As demonstrated in Sect.~\ref{subsec:rot_subcl}, the rotators form a particular sub-sample of objects even among the
sources in our main sample. In this section we investigate whether there are any differences in the gamma-ray properties
between these two sub-classes. To this end, we extract the variability indices and we calculate luminosities in the
gamma-ray band ($100\,{\rm MeV} \le E \le 100\,{\rm GeV}$) for blazars of our main sample using the data from the 3FGL catalogue \citep{Acero2015}.
The cumulative distribution functions (CDF) of these quantities for rotators and non-rotators are shown in
Fig.~\ref{fig:gamma_lum}. According to the two-sample K-S test there is a strong indication that both luminosity
($p\text{-value} = 0.02$) and variability ($p\text{-value} = 0.01$) are higher for the blazars that exhibited rotations.

This is presumably caused by the dominance of LSP sources among rotators found in the previous section, since LSP blazars
tend to have higher gamma-ray luminosities than HSP sources \citep{Ackermann2015}.
High variability indices in the gamma-ray band are characteristic of sources that are both luminous and variable
\citep{Ackermann2015}. Therefore the difference in the variability indices is also explained by the dominance of LSP
blazars among the rotators.

\section{Discussion and conclusions} \label{sec:conclusion}

We have presented a set of EVPA rotations detected by {\em RoboPol} during the 2015 observing season. After three
years of operation we have detected 40 EVPA rotations, and thereby more than tripled the list of known events of
this type.

Our monitoring sample was constructed on the basis of statistically robust and bias-free criteria. It included both
gamma-ray--loud and gamma-ray--quiet blazars that were monitored with equal cadence. This allowed us to perform
statistical studies of the frequency of EVPA rotations in blazars for the first time.

We have shown that the frequency of rotations varies significantly among blazars. None of the control sample blazars
displayed a rotation during the monitoring period. Moreover, the EVPA rotations occur with significantly different
frequency in different blazars in the main sample. There is a subset of blazars that show the events much more frequently than
others. This result is consistent with our analysis in Paper~I, where we showed that rotators have higher EVPA
variability than non-rotators even outside the rotating periods.

This is a major result of the {\em RoboPol} project: only a fraction of blazars ($\sim$ 28\% of sources in both samples)
exhibit EVPA rotations with rates $\le 20$ deg d$^{-1}$ in the optical band, with an average frequency of 1/232\,d$^{-1}$
(in the observer frame). The remaining $\sim$ 72\% of sources did not show any rotations. If they do exhibit rotations,
this should happen with a frequency less than $\sim$ 1/3230\,d$^{-1}$.

The analysis of Sect.~\ref{sec:reasons} shows that the difference in the frequencies of EVPA rotations cannot be explained
either by the difference in the EVPA measurement uncertainties or by differences in redshifts and/or Doppler factors
among the blazars. This result should be confirmed using a larger number of objects with known $\delta$. Only a small fraction
of blazars in our monitoring sample have Doppler factor estimates available. The ongoing analysis of variability in the
radio band will allow us to increase the sample of blazars with known Doppler factors and allow to verify our results
with better statistics.

The tendency for EVPA rotations to occur in LSP blazars found in Sect.~\ref{sec:rot_classes} can be
explained in the same way as higher variability of LSP sources in the total optical flux. It has been shown by \cite{Hovatta2014}
that LSP blazars are more variable than HSP in the optical band. This was attributed to the fact that, in the optical band,
LSP sources are observed near their electron energy peak, which causes stronger variations of the emission compared to HSP sources,
where the lower energy electrons cool down slowly and produce mild variability. For the same reason, the polarized flux
density as well as the EVPA must be more variable in LSP sources compared to HSP when observed in the optical band. If
this interpretation is correct, then HSP blazars must exhibit EVPA rotations more frequently at higher frequencies (UV and
X-ray bands). The dependence of the optical EVPA behaviour on the synchrotron peak position is also reported
in two other papers based on RoboPol data. \cite{Angelakis2016} have shown that the EVPA in HSP sources centers around a
preferred direction, while in LSP blazars it follows a more uniform distributions. \cite{Hovatta2016} have shown that the
scatter in the $Q$-$U$ plane is smaller for HSP blazars than for ISP. This also indicates that the polarization plane direction
is more stable in HSP sources.

We also found that the rotators seem to be more luminous and more variable in the gamma-ray band than non-rotators. This
difference can also be explained by the tendency of the EVPA rotations to occur in LSP sources. These sources have higher
luminosities on average than ISP and HSP in the 3FGL because of an instrumental selection effect. The same reason can also
explain the increase of their variability indices \citep{Ackermann2015}. For this reason, the optical polarimetry monitoring
programmes that select their observing samples on the basis of high variability in the gamma-ray band will observe EVPA
rotations more frequently than among blazars on average.

The 180$^{\circ}$ EVPA ambiguity sets a fundamental limitation on the rate of EVPA rotations that can be detected
under a given cadence of observations. So far we have been able to study rotations with rates $\le 20$\,deg\,d$^{-1}$. There
was only one rotation with a rate $\sim$ 50\,deg\,d$^{-1}$ detected by {\em RoboPol}. However, there is an indication in the
{\em RoboPol} data as well as in the literature that fast EVPA rotations with rates 60 -- 130\,deg\,d$^{-1}$ do occur in blazars
\citep[e.g.][]{Larionov2013}. We plan to extend our studies to higher rotation rates by increasing our cadence for
future monitoring.

\section*{Acknowledgements}
The RoboPol project is a collaboration between the University of Crete/FORTH in Greece, Caltech in the USA,
MPIfR in Germany, IUCAA in India and Toru\'{n} Centre for Astronomy in Poland.
The U. of Crete/FORTH group acknowledges support by the ``RoboPol'' project, which is
co-funded by the European Social Fund (ESF) and Greek National Resources, and by the European
Comission Seventh Framework Programme (FP7) through grants PCIG10-GA-2011-304001 ``JetPop'' and
PIRSES-GA-2012-31578 ``EuroCal''.
This research was supported in part by NASA grant NNX11A043G and NSF grant AST-1109911, and by the
Polish National Science Centre, grant number 2011/01/B/ST9/04618.
DB acknowledges support from the St. Petersburg University research grant 6.38.335.2015.
KT acknowledges support by the European Commission Seventh Framework Programme (FP7) through
the Marie Curie Career Integration Grant PCIG-GA-2011-293531 ``SFOnset''.
MB acknowledges support from NASA Headquarters under the NASA Earth and Space Science Fellowship Program, grant NNX14AQ07H.
TH was supported by the Academy of Finland project number 267324.
IM and SK are supported for this research through a stipend from the International Max Planck
Research School (IMPRS) for Astronomy and Astrophysics at the Universities of Bonn and Cologne.




\bibliographystyle{mnras}
\bibliography{bibliography_manual}

\appendix

\section{Redshifts, Doppler factors and synchrotron peak classes} \label{ap:a}

\begin{landscape}
\begin{table}
\centering
\scriptsize
\caption{}
\label{tab:app}
  \begin{tabular}{lccccclccccc} 
  \hline
 Blazar ID     &   Survey         &  $z$          & $\delta^1$& synch.     &     $z$ ref.                  & Blazar ID       &   Survey         &  $z$          & $\delta^1$    & synch.     &     $z$ ref.           \\
 RBPL...       &   name           &               &           & peak class &                               & RBPL...         &   name           &               &               & peak class &   \\
 \hline
J0017+8135*    &   S5 0014+81     &    3.366      &     -     &    LSP     & \citep{1994ApJ...436..678O}   &  J1800+3848*    &     S4 1758+38   &    2.092      &     -         &    HSP     & \citep{1994AAS..103..349S} \\
J0045+2127$^r$ &  GB6J0045+2127   &      -        &     -     &    HSP     &      -                        &  J1800+7828$^r$ &   S5 1803+784    &    0.684      &    12.2       &    LSP     & \citep{2001AJ....122..565R} \\
J0114+1325     &  GB6J0114+1325   &  $0.583^\dag$ &     -     &    ISP     & \citep{2014ApJ...784..151S}   &  J1806+6949$^r$ &      3C 371      &    0.051      &    1.1        &    ISP     & \citep{1992AAS...96..389d} \\
J0136+4751$^r$ &     OC 457       &    0.859      &   20.7    &    LSP     & \citep{1987ApJS...63....1H}   &  J1809+2041$^r$ &   RXJ1809.3+2041 &      -        &     -         &    HSP     &     -                      \\
J0211+1051     &  MG1J021114+1051 &   $0.2^\dag$  &     -     &    ISP     & \citep{2010ApJ...712...14M}   &  J1813+3144     &   B2 1811+31     &    0.117      &     -         &    ISP     & \citep{1991ApJ...378...77G} \\
J0217+0837     &     ZS0214+083   &    0.085      &     -     &    LSP     & \citep{Shaw2013}              &  J1835+3241*    &     4C 32.55     &   0.0579      &     -         &     -      & \citep{1996AJ....112.1803M} \\
J0259+0747$^r$ &   PKS 0256+075   &    0.893      &     -     &    LSP     & \citep{1993MNRAS.264..298M}   &  J1836+3136$^r$ &   RXJ1836.2+3136 &      -        &     -         &    ISP     &     -                      \\
J0303$-$2407   &   PKS 0301$-$243 &    0.2657     &     -     &    HSP     & \citep{2014AA...565A..12P}    &  J1838+4802     &   GB6J1838+4802  &   $0.3^\dag$  &     -         &    HSP     & \citep{2003AA...400...95N} \\
J0405$-$1308   &   PKS 0403$-$13  &    0.5706     &     -     &    ISP     & \citep{1996ApJS..104...37M}   &  J1841+3218     &   RXJ1841.7+3218 &      -        &     -         &    HSP     &     -                      \\
J0423$-$0120   &   PKS 0420$-$01  &    0.9161     &   19.9    &    LSP     & \citep{2009MNRAS.399..683J}   &  J1854+7351*    &    S5 1856+73    &    0.461      &     -         &    LSP     & \citep{1997MNRAS.290..380H} \\
J0642+6758*    &    S4 0636+68    &    3.18       &     -     &    LSP     & \citep{1994ApJ...436..678O}   &  J1903+5540     &    TXS1902+556   &   $0.58^\dag$ &     -         &    ISP     & \citep{2010ApJ...712...14M} \\
J0825+6157*    &    TXS0821+621   &    0.542      &     -     &    LSP     & \citep{1987ApJS...63....1H}   &  J1927+6117$^r$ &   S4 1926+61     &   $0.54^\dag$ &     -         &    LSP     & \citep{2010ApJ...712...14M} \\
J0841+7053     &     4C 71.07     &    2.172      &   16.3    &    LSP     & \citep{1993AAS..100..395S}    &  J1927+7358*    &   4C 73.18       &    0.3021     &    1.9        &    LSP     & \citep{1996ApJS..104...37M} \\
J0848+6606     &   GB6J0848+6605  &      -        &     -     &    ISP     &      -                        &  J1955+5131*    &    S4 1954+51    &    1.23       &    7.4        &    LSP     & \citep{1994AAS..103..349S} \\
J0854+5757*    &     4C 58.17     &    1.3192     &     -     &    ISP     & \citep{2010MNRAS.405.2302H}   &  J1959+6508     &   1ES 1959+650   &    0.047      &     -         &    HSP     & \citep{1993ApJ...412..541S} \\
J0957+5522     &     4C 55.17     &    0.8996     &     -     &    ISP     & \citep{2010MNRAS.405.2302H}   &  J2005+7752     &    S5 2007+77    &    0.342      &    7.9        &    LSP     & \citep{1989AAS...80..103S} \\
J0958+6533     &   S4 0954+65     &  $0.368^\dag$ &    6.2    &    LSP     & \citep{1992ApJ...398..454W}   &  J2015$-$0137   &   PKS 2012$-$017 &      -        &     -         &    ISP     &     -                       \\
J1037+5711$^r$ &   GB6J1037+5711  & $0.8304^\dag$ &     -     &    ISP     & \citep{2009AJ....137.3884R}   &  J2016$-$0903   &  PMNJ2016$-$0903 &    0.367      &     -         &    ISP     & \citep{2011ApJ...743..171A} \\
J1048+7143$^r$ &   S5 1044+71     &    1.150      &     -     &    LSP     & \citep{1995ApJS...98....1P}   &  J2016+1632*    &    TXS 2013+163  &      -        &     -         &     -      & - \\
J1058+5628     &   TXS1055+567    &  $0.143^\dag$ &     -     &    HSP     & \citep{Shaw2013}              &  J2022+7611$^r$ &    S5 2023+760   &    0.594      &     -         &    ISP     & \citep{2011ApJ...743..171A} \\
J1203+6031     &   SBS1200+608    &    0.0656     &     -     &    ISP     & \citep{1998ApJ...494...47F}   &  J2024+1718*    &  GB6 J2024+1718  &    1.05       &    5.8        &    LSP     & \citep{1996cqan.book.....V} \\
J1248+5820     &   PG 1246+586    &    0.8474     &     -     &    ISP     & \citep{2005AJ....129.1755A}   &  J2030$-$0622   &   TXS2027$-$065  &    0.667      &     -         &    LSP     & \citep{Ackermann2015} \\
J1512$-$0905$^r$&  PKS 1510$-$089 &    0.360      &    16.7   &    LSP     & \citep{1990PASP..102.1235T}   &  J2033+2146*    &      4C 21.55    &   0.1735      &     -         &     -      & \citep{2007ApJ...664...64I} \\
J1542+6129     &   GB6J1542+6129  &  $0.117^\dag$ &     -     &    ISP     & \citep{2011ApJ...740...98M}   &  J2039$-$1046   &    TXS2036$-$109 &   $1.05^\dag$ &     -         &    LSP     & \citep{2013AJ....146..127S} \\
J1551+5806*    &   GB6 J1551+5806 &    1.324      &     -     &    ISP     & \citep{1999MNRAS.307..149S}   &  J2042+7508*    &     4C 74.26     &    0.104      &     -         &     -      & \citep{1996AAS..115....1S} \\
J1553+1256     &   PKS 1551+130   &  $1.308^\dag$ &     -     &    ISP     & \citep{2014MNRAS.438.3058R}   &  J2131$-$0915   &  BZB J2131$-$0915&    0.449      &     -         &    HSP     & \citep{2005AA...434..385G} \\
J1555+1111$^r$ &   PG 1553+113    &  $0.360^\dag$ &     -     &    HSP     & \citep{2011ApJS..194...29R}   &  J2143+1743     &    OX 169        &    0.2107     & $8.8\pm1.8^2$ &    ISP     & \citep{2009ApJS..184..398H} \\
J1558+5625$^r$ &   TXS1557+565    &  $0.300^\dag$ &     -     &    ISP     & \citep{1998ApJ...494...47F}   &  J2148+0657     &     4C 6.69      &    0.999      &   15.6        &    LSP     & \citep{1991ApJ...382..433S} \\
J1603+5730*    &    4C 57.27      &  $2.4084^\dag$&     -     &    ISP     & \citep{1991ApJS...77..203J}   &  J2149+0322     &    PKS B2147+031 &      -        &     -         &    ISP     &     -                      \\
J1604+5714     &   GB6J1604+5714  &    0.720      &     -     &    ISP     & \citep{1998ApJ...494...47F}   &  J2150$-$1410   &   TXS2147$-$144  &    0.229      &     -         &    HSP     & \citep{1998AN....319..347F} \\
J1607+1551     &    4C 15.54      & $0.4965^\dag$ &     -     &    LSP     & \citep{2008ApJS..175..297A}   &  J2202+4216$^r$ &     BL Lac       &    0.0686     &   7.3         &    LSP     & \citep{1995ApJ...452L...5V} \\
J1624+5652*    &    7C 1623+5659  &  $0.415^\dag$ &     -     &     -      & \citep{2009ApJS..180...67R}   &  J2225$-$0457   &      3C 446      &    1.404      &   16.0        &    LSP     & \citep{1983MNRAS.205..793W} \\
J1635+3808$^r$ &    4C 38.41      &    1.8131     &    21.5   &    LSP     & \citep{2010MNRAS.405.2302H}   &  J2232+1143$^r$ &     CTA 102      &    1.037      &   15.6        &    LSP     & \citep{1965ApJ...141.1295S} \\
J1638+5720*    &    S4 1637+57    &    0.7506     &    14.0   &    LSP     & \citep{1996ApJS..104...37M}   &  J2243+2021$^r$ &   RGBJ2243+203   &   $0.39^\dag$ &     -         &    HSP     & \citep{2010ApJ...712...14M} \\
J1642+3948     &    3C 345        &    0.5928     &    7.8    &    LSP     & \citep{1996ApJS..104...37M}   &  J2251+4030     & BZB J2251+4030   &    0.229      &     -         &    ISP     & \citep{2011ApJ...743..171A} \\
J1653+3945     &    Mkn 501       &    0.0337     &     -     &    HSP     & \citep{1993AAS...98..393S}    &  J2253+1608$^r$ &     3C 454.3     &    0.859      &   33.2        &    LSP     & \citep{1991MNRAS.250..414J} \\
J1725+1152     &    1H 1720+117   & $0.018^\dag$  &     -     &    HSP     & \citep{1993ApJS...85..111C}   &  J2311+3425$^r$ &  B2 2308+34      &    1.817      &     -         &    LSP     & \citep{1976ApJS...31..143W} \\
J1748+7005$^r$ &    S4 1749+70    &    0.770      &     -     &    LSP     & \citep{1992ApJ...396..469H}   &  J2334+0736     &    TXS2331+073   &    0.401      &     -         &    LSP     & \citep{2005ApJ...626...95S} \\
J1751+0939$^r$ &     OT 081       &    0.320      &    12.0   &    LSP     & \citep{1993AAS...98..393S}    &  J2340+8015     &  BZB J2340+8015  &    0.274      &     -         &    HSP     & \citep{2002ApJ...566..181C} \\
J1754+3212     &   RXJ1754.1+3212 &  $1.09^\dag$  &     -     &    ISP     & \citep{2013AJ....146..127S}   &                 &                  &               &               &            &                             \\
\hline
\multicolumn{6}{l}{* - control sample source; ${}^r$ - rotator; ${}^\dag$ - non-spectroscopic $z$;}\\
\multicolumn{6}{l}{${}^1$ from \cite{Hovatta2009} unless other reference is specified; ${}^2$ from Liodakis et al. (in prep.).}
\end{tabular}
\end{table}
\end{landscape}

\bsp	
\label{lastpage}
\end{document}